\documentclass{article}
\usepackage[english]{babel}
\usepackage{amsmath,amssymb,bbm,theorem}
\usepackage{tikz}

\catcode`\>=\active \def>{
\fontencoding{T1}\selectfont\symbol{62}\fontencoding{\encodingdefault}}
\newcommand{\assign}{:=}
\newcommand{\nocomma}{}
\newcommand{\noplus}{}
\newcommand{\precprec}{\prec\!\!\!\prec}
\newcommand{\tmdummy}{$\mbox{}$}
\newcommand{\tmem}[1]{{\em #1\/}}
\newcommand{\tmmathbf}[1]{\ensuremath{\boldsymbol{#1}}}
\newcommand{\tmop}[1]{\ensuremath{\operatorname{#1}}}
\newcommand{\tmscript}[1]{\text{\scriptsize{$#1$}}}
\newcommand{\tmstrong}[1]{\textbf{#1}}
\newtheorem{definition}{Definition}
{\theorembodyfont{\rmfamily}\newtheorem{example}{Example}}
\newtheorem{lemma}{Lemma}
\newtheorem{proposition}{Proposition}
{\theorembodyfont{\rmfamily}\newtheorem{remark}{Remark}}
\newtheorem{theorem}{Theorem}

 \newcommand{\Sabplusc}{\begin{tikzpicture}[scale=0.7,baseline=5pt]
  \node (r) at (0,0) {$\alpha$};
  \node (a) at (0,1) {${\beta} +{\gamma}$};
    \draw (r) --(a) ;
 \end{tikzpicture}}
 \newcommand{\Sabc}{\begin{tikzpicture}[scale=0.7,baseline=5pt]
  \node (a) at (0,0) {${\alpha}$};
  \node (b) at (0,1) {${\beta}$};
  \node (c) at (0,2) {${\gamma}$};
    \draw (a) --(b) ;
    \draw (b) --(c) ;
 \end{tikzpicture}}
\newcommand{\Sacb}{\begin{tikzpicture}[scale=0.7,baseline=5pt]
  \node (a) at (0,0) {${\alpha}$};
  \node (c) at (0,1) {${\gamma}$};
  \node (b) at (0,2) {${\beta}$};
    \draw (a) --(c) ;
    \draw (c) --(b) ;
 \end{tikzpicture}}
 \newcommand{\Fbunbdeux}{\begin{tikzpicture}[scale=0.7,baseline=-2pt]
  \node (r) at (0,0) {$B_1$};
  \node (a) at (1,0) {$B_2$};
 \end{tikzpicture}}
 \newcommand{\Sbunbdeux}{\begin{tikzpicture}[scale=0.7,baseline=5pt]
  \node (r) at (0,0) {$B_1$};
  \node (a) at (0,1) {$B_2$};
    \draw (r) --(a) ;
 \end{tikzpicture}}
 \newcommand{\Sbunbtrois}{\begin{tikzpicture}[scale=0.7,baseline=5pt]
  \node (r) at (0,0) {$B_1$};
  \node (a) at (0,1) {$B_3$};
    \draw (r) --(a) ;
 \end{tikzpicture}}
 \newcommand{\BaFbc}{
 \begin{tikzpicture}[scale=0.7,baseline=5pt]
  \node (r) at (0,0) {${\alpha}$};
  \node (a) at (-0.5,1) {${\beta}$};
  \node (b) at (0.5,1) {${\gamma}$};
  \draw (r) --(a) ;
  \draw (r) --(b) ;
\end{tikzpicture}}
\newcommand{\Sbunbdeuxbtrois}{
\begin{tikzpicture}[scale=0.7,baseline=5pt]
  \node (r) at (0,0) {$B_1$};
  \node (a) at (0,1) {$B_2$};
  \node (b) at (0,2) {$B_3$};
    \draw (r) --(a) ;
    \draw (a) --(b) ;
 \end{tikzpicture}}
  \newcommand{\Sbdeuxbtrois}{\begin{tikzpicture}[scale=0.7,baseline=5pt]
  \node (r) at (0,0) {$B_2$};
  \node (a) at (0,1) {$B_3$};
    \draw (r) --(a) ;
 \end{tikzpicture}}
 \newcommand{\Fbunbdeuxbtrois}{\begin{tikzpicture}[scale=0.7,baseline=-2pt]
  \node (r) at (0,0) {$B_1$};
  \node (a) at (1,0) {$B_2$};
  \node (a) at (2,0) {$B_3$};
 \end{tikzpicture}}
 \newcommand{\BbunFbdeuxbtrois}{
 \begin{tikzpicture}[scale=0.7,baseline=5pt]
  \node (r) at (0,0) {$B_1$};
  \node (a) at (-0.5,1) {$B_2$};
  \node (b) at (0.5,1) {$B_3$};
  \draw (r) --(a) ;
  \draw (r) --(b) ;
\end{tikzpicture}}

\begin{document}

\title{Ecalle's paralogarithmic resurgence monomials and effective synthesis}

\author{Fr{\'e}d{\'e}ric Fauvet}

\maketitle

\begin{abstract}
  Paralogarithms constitute a family of special functions, which are some
  generalizations of hyperlogarithms. They have been introduced by Jean Ecalle
  in the context of the classification of complex analytic dynamical systems
  with irregular singularities, to solve the so--called ``synthesis problem''
  in an effective and very general way.
  
  We describe the formalism of resurgence monomials, introduce the
  paralogarithmic family and present the effective synthesis with
  paralogarithmic monomials, of analytic vector fields having a saddle--node
  singularity, following Ecalle.
\end{abstract}

\section{Introduction}

Resurgent functions and alien calculus were introduced by Jean Ecalle more
than 40 years ago for classifying singularities of dynamical systems with
analytic data. Divergent series appearing in the solutions of these complex
analytic dynamical systems at irregular singularities, {\tmem{when expressed
in a suitable variable}} $z \sim \infty$ can quite systematically be treated
by the Borel--Laplace summation formalism (see the next section). The Borel
transform of these series have isolated singularities and display properties
of ``self--reproduction'' at their singular points ({\cite{e2,e3}}, which are
at the origin of the very name resurgence; families of so--called alien
operators were defined by Ecalle to characterize this resurgent behaviour.

Alien calculus has made it possible to solve difficult problems of moduli in
an effective manner, which was the ultimate goal; many consequences for local
dynamics can then easily be derived thanks to the precise and effective
description of the moduli spaces obtained with the help of alien derivations
({\cite{e3,e_snag}}). Beyond the original field of applications, the
phenomenon of resurgence had been described in the early 80's for perturbation
series in quantum mechanics and quantum field theory; more recently, there has
been an explosion of works in theoretical Physics involved with resurgence
(see e.g.  \cite{gmp,du,cms,bor_dunne} and notably {\cite{abs2019}} and the references therein) -- and significant connections with ``wall--crossing" (\cite{gmn,ks}).

The classification of analytic saddle--nodes, with a complete description of
the corresponding moduli spaces, had been a major breakthrough in the
understanding of the irregular singularities of complex analytic dynamical
systems. It has been obtained around 1980, independently by J. Ecalle, with
resurgent functions, and by J. Martinet and J.--P. Ramis
({\cite{martinet_ramis}}), who pioneered a ``cohomological approach'' to study
the Stokes phenomenon.

Saddle--node is a traditional terminology to designate singularities of
differential equations which display in the real phase--space a set of
trajectories with a saddle behaviour on one side and a node behaviour on the
other. The formal classification of these dynamical systems and their
generalizations, in 2 complex dimensions, had been known for a long time and
there were partial results regarding the transformations that normalize these
germs of vector fields.

Such formal normalizations correspond to analytic functions in the critical
variable, but only in sectorial domains stemming from the singular point under
study: the formal series involved are generically divergent. This divergence
was characterized by Martinet--Ramis by a delicate study of the differences of
these analytic sectorial functions on overlapping sectors and of the
corresponding Stokes phenomenon; in their approach, some non--abelian first
cohomology space characterizes the moduli. 

For the simplest formal normal form, a point in the moduli space amounts to the data of a pair of Stokes diffeomorphisms $S_+,S_-$ in the respective singular directions (cf below) $\mathbbm{R}_{>}$ and $\mathbbm{R}_{<}$; the diffeo. $S_-$ being a translation by a constant and $S_+$ being any analytic tangent to identity germ of diffeo.
To solve the inverse problem, notably, Martinet--Ramis's scheme performed the synthesis  in a non--effective way of a vector field  $X$ eventually yielding the given pair $S_+,S_-$ first in the $\cal{C}^\infty$ category and then using the crucial Newlander--Nirenberg integrability theorem to obtain analyticity ({\cite{martinet_ramis}). 

On his side, Ecalle solved the problem by the use of alien calculus and had in
particular presented a solution to the inverse problem (``synthesis''), for moduli of germs of tangent to identity diffeomorphisms in dimension $1$, early
on in vol. 2 of {\cite{e2}}, yet in a non constructive way; more recently, however, he had described in
{\cite{e_twist}} a family of new special functions, the
so--called {\tmem{paralogarithmic resurgence monomials}}, that can be used in
an effective way to synthesize {\tmem{in full generality}}
analytic objects which have prescribed moduli expressed in the resurgent
formalism.

The aim of this text is to present these beautiful objects, describe in the
framework of Ecalle's theories this systematic and constructive solution to
the synthesis problem, focusing on the example of the saddle--node singularity
-- and restraining to the simplest formal normal form.

We indicate now how the present text is organized. To make the article
self--contained, we begin by giving a brief introduction to resurgent
functions and alien derivations. Next, we present all the necessary background
on {\tmem{mould calculus}} which is required to formulate the scheme for the
resurgent treatment of the synthesis problem and to solve it, using notably
the crucial apparatus of {\tmem{arborification--coarborification}}.

We recall after that the results pertaining to the resurgence properties of
the saddle-node (``the direct problem'') and then move to the ``inverse
problem'', with a presentation of J. Ecalle's way to solve it. We introduce
Ecalle's paralogarithmic resurgence monomials and give schemes of proofs for their
properties, to eventually reach a complete and explicit solution of the
synthesis problem.

\begin{remark}
  {\tmem{No claim of originality}} is made concerning the main results
  contained in the present article, which can be viewed as
  {\tmstrong{{\tmem{an invitation to read Ecalle's foundational texts, in
  particular {\cite{e_twist}}}}}}. Resurgent functions, alien calculus,
  moulds, arborification, the problematics of resurgent monomials, the
  resurgent treatment of the problem of synthesis, paralogarithmic monomials
  themselves and the schemes to prove their essential properties are all due
  to Ecalle. 
\end{remark}

{\noindent}{ \bf Acknowledgments } We are very grateful to Jean Ecalle for
ever enlightening explanations (remaining imprecisions, if any, remain of course ours); to Ricardo Schiappa and David Sauzin for many discussions and suggestions, notably about the
references {\cite{cv, gmn}} by Cecotti--Vafa and Gaiotto-Moore-Neitzke and in particular the striking occurences in these
works of integral formulas to tackle linear inverse problems, which are quite close to the ones defining
paralogarithms; to Dominique Manchon for many exchanges on this subject and remarks on preliminary versions of the text; to Jasmin Raissy  and the organizers of the special program ``Resurgent Asymptotics in Physics
and Mathematics'', at KITP in 2017, for invitations to lecture on these topics, in Toulouse and the Kavli Institute respectively.{\hspace*{\fill}}{\medskip}

\section{Resurgent functions and alien differential calculus}

\subsection{Basic facts about resurgent functions}

We briefly recall basic definitions and properties of resurgent functions,
adapted to the context of saddle--node singularities which we shall study in
detail below.

Most resurgent functions of natural origin (say, formal expansions appearing
in the general solution of some functional equation or expansions in some
critical parameter in Physics) appear as Borel transforms with respect to some
critical variable $z \sim \infty$, where the Borel transform $\mathcal{B}$ of
a formal series $\tilde{\varphi} (z) = \sum_{n \geqslant 0} c_n z^{- (n + 1)}$
is:
\[ \mathcal{B} (\tilde{\varphi}) \assign \hat{\varphi} : \zeta \longrightarrow
   \sum_{n \geqslant 0} c_n  \frac{\zeta^n}{n!} \]
As a rule, the presence of exponentials of the type $e^{a z}$ ($a \in
\mathbbm{C}$) in formal solutions at an irregular singularity of a given
dynamical system is concomitant with the Gevrey--1 type of divergence of the
formal series implied.

\begin{definition}
  A formal series $\sum_{n \geqslant 0} a_n t^{n + 1}$ is called Gevrey--$1$
  iff $\exists C \nocomma, M$ >0 such that
  \[ | a_n | \leqslant C M^n n! \quad, \forall n \geqslant 0 \]
  
\end{definition}

Any series $\tilde{\varphi} (z) = \sum_{n \geqslant 0} c_n z^{- (n + 1)}$
which is Gevrey--1 has a Borel transform $\hat{\varphi} (\zeta) = \sum_{n
\geqslant 0} c_n  \frac{\zeta^n}{n!}$ which belongs to $\mathbbm{C} \{ \zeta
\}$ and it is a general fact that the $\hat{\varphi} (\zeta)$ corresponding to
formal series involved in the solutions at the singularity under study can be
analytically continued, with isolated singularities $\omega$ in the
$\zeta$--plane. There are operators with good algebraic properties which make
it possible to characterize the singular behaviour at the points $\omega$,
notably the so--called alien derivations $\Delta_{\omega}$ defined below.

If $\hat{\varphi} (\zeta)$ has at most exponential growth in the direction
$\mathbbm{R}_{>}$ (meaning that $\exists H, K > 0$ such that $\forall \zeta
\in \mathbbm{R}_{>}$, $| \hat{\varphi} (\zeta) | \leqslant H e^{K \zeta}$), we
can perform a Laplace transform of this function and get a function:
\[ \varphi (z) \assign \mathcal{L} (\hat{\varphi}) (z) = \int_0^{\infty} e^{-
   \zeta z}  \hat{\varphi} (\zeta) d \zeta \]
This function $\varphi (z)$ is analytic in the half plane $\tmop{Re} (z) > K$;
more generally, for any direction $d_{\theta} = e^{i \theta} \mathbbm{R}_{>}$
from the origin in the $\zeta$--plane, a Laplace integration in the direction
$d_{\theta}$
\[ \mathcal{L}_{\theta} (\hat{\varphi}) (z) = \int_0^{e^{i \theta} \infty}
   e^{- \zeta z}  \hat{\varphi} (\zeta) d \zeta \]
yields a function $\varphi (z)$ which is analytic in some half plane bisected
by the conjugate direction $\bar{d} \assign e^{- i \theta} \mathbbm{R}^{>}$
and this scheme is licit whenever $\hat{\varphi} (\zeta)$ can be analytically
continued with exponential growth at $\infty$ along $d_{\theta}$.

The Borel transform $\mathcal{B}$ is a morphism of differential algebras, where the
product in the $\zeta$ plane is the following convolution (and the derivation
wrt this convolution is multiplication by $- \zeta$):
\[ \hat{\varphi} \ast \hat{\psi}  (\zeta) = \int_0^{\zeta} \hat{\varphi} (s)
   \hat{\psi} (\zeta - s) d s \quad (\zeta \thicksim 0) \]
Resurgent functions appear in practice in 3 ``models'' :
\begin{enumerate}
  \item The formal model, involving formal series in $z^{- 1}$
  
  \item The geometric model(s), involving holomorphic functions in sector(s)
  
  \item The Borel plane, involving functions with isolated singularities and
  tame behaviour at $\infty$
\end{enumerate}
In fact, we shall very soon need to consider germs, defined on the Riemann
surface of the logarithm $\mathbbm{C}_{\infty}$, which are a priori also
singular at the origin; these 3 models and the morphisms that relate them will
be accordingly enriched ({\cite{e3,e_snag,e_twist}}).

Let us focus on the Borel plane and consider a function $\psi (\zeta)$ \ which
is defined in particular for any $\zeta$ in a universal cover
$\widetilde{D_{\ast}}$ of a pointed disk $D_{\ast} \assign D (\omega, \rho) -
\{ \omega \}$ centered at a given complex number $\omega_{}$, which is thus a
priori a singular point of $\psi (\zeta)$; the singularity of $\psi (\zeta)$
at $\omega$ can be characterized by the class of $\psi (\zeta)$, mod. analytic
germs at $\omega$, for $\zeta$ on $\widetilde{D_{\ast}}$.

We now proceed to define resurgent functions, at a level of
generality which is sufficient for the present text (we notably refer to
{\cite{e3}} for more general cases).

\begin{definition}
  A function, defined at the origin of $\mathbbm{C}_{\infty}$ (meaning: for any $\zeta$ in a universal cover
$\widetilde{D_{\ast}}$ of a pointed disk $D_{\ast} \assign D (0, \rho) -
\{0 \}$), is called endlessly continuable iff it can be
  analytically continued, with isolated singularities. The set of germs of
  functions which are endlessly continuable is denoted by $\mathcal{P}$.
\end{definition}

The set $\mathcal{P}$ has a natural structure of vector space; it encodes any
type of ramified singularity at the origin of $\mathbbm{C}$ and, by discarding
the functions which are regular at the origin of $\mathbbm{C}$ we arrive at
the following:

\begin{definition}
  Let $\mathcal{R}=\mathcal{P}/\mathcal{P}_1$ the quotient of $\mathcal{P}$ by
  the subspace $\mathcal{P}_1$ of regular germs ($\mathcal{P}_1 =\mathcal{P}
  \cap \mathbbm{C} \{ \zeta \}$).
  
  An element of $\mathcal{R}$ will be denoted by $\varphi^{\nabla}$ (or simply
  $\varphi$ for short, keeping in mind that $\varphi$ designates a class of
  functions) and any representative of such an element will be called a major
  of $\varphi^{\nabla}$and denoted by $\check{\varphi}$. For any major
  $\check{\varphi}$ of a given element $\varphi^{\nabla}$in $\mathcal{R}$, the
  function $\hat{\varphi}$ defined for $\zeta \sim 0$ by
  \[ \hat{\varphi} (\zeta) = (1 - R) \check{\varphi} (\zeta) = \check{\varphi}
     (\zeta) - \check{\varphi} (\zeta \tmmathbf{e}^{- 2 \pi i}) \qquad (R
     \tmop{is} \tmop{the} \tmop{rotation} \tmop{of} - 2 \pi, \tmop{on}
     \mathbbm{C}_{\infty}) \]
  does not depend on the choice of the major; it is called the minor of
  $\varphi^{\nabla}$
\end{definition}

\begin{remark}
  One thing can be highlighted from the
  start: as defined above, resurgent functions are not functions -- they are classes of
  functions -- and {\tmem{stricto sensu}} there is nothing resurgent in them,
  either.
  It is {\tmem{in the applications}} that the phenomenon of resurgence will be
  displayed and the vast majority of divergent formal series of natural origin,
  appearing in solutions to dynamical systems at singularities, or in various
  physical theories, show some form of resurgent behaviour: at the
  singularities in the Borel plane of a given solution, some other solution
  resurges, in a shape that depends upon the problem under study. 
\end{remark}

\begin{example}
  To any analytic regular germ $\varphi (\zeta)$ at the origin of
  $\mathbbm{C}$ we can associate a germ on $\mathbbm{C}_{\infty}$ which
  defines a logarithmic singularity, namely $\varphi^{\nabla}$ defined by the
  major:
  \[ \check{\varphi} (\zeta) = \frac{1}{2 \pi i} \log (\zeta) \varphi (\zeta)
  \]
  The minor of $\varphi^{\nabla}$ is the germ $\varphi (\zeta)$; by abuse of
  language, in this case it is harmless to speak there of ``the resurgent
  function $\hat{\varphi}  (\zeta)$'' (or simply the function $\varphi  (\zeta)$)
\end{example}

Ecalle's formalism of majors and minors makes it possible to deal with any
isolated singularity in the Borel plane and to eventually define alien
operators for very general spaces of resurgent functions, but for our purposes
in the present text, we shall mostly work with resurgent functions with
analytic continuations which have simple singularities, according to the
following:

\begin{definition}
  We say that the analytic continuation along a path $\gamma$ from the origin
  to $\omega \in \mathbbm{C}$ of a function $\hat{\varphi} (\zeta) \in
  \mathbbm{C} \{ \zeta \}$ has a {\tmem{simple singularity}} at the point
  $\omega$ if and only if there exist 2 analytic functions $a (\zeta)$, $b
  (\zeta)$ at $0$ and a constant $\alpha_{\omega}$ such that
  \[ \hat{\varphi} (\omega + \zeta) = \frac{\alpha_{\omega}}{2 \pi i \zeta} +
     \frac{1}{2 \pi i}  (\log \zeta) a (\zeta) + b (\zeta) \]
  The singularity at $\omega$ is thus characterized by the data
  ($\alpha_{\omega}, a (\zeta)$) and the minor is $a (\zeta)$
\end{definition}

For such functions, the minor is equal to the germ $a (\zeta)$ and the
singularity, up to the simple pole $\frac{\alpha_{\omega}}{2 \pi i \zeta}$, is
characterized by the minor.

\begin{remark}
  In the present text, the resurgent functions we deal with have minors which
  are uniform at the origin; it that case, it is harmless to keep the same
  letter $\zeta$ for germs at the origin of $\mathbbm{C}$ or
  $\mathbbm{C}_{\infty}$. 
\end{remark}

There is a convolution product, defined at the level of majors, that boils down,
for logarithmic singularities, to the formula given above; we shall not need
the general convolution product, for the saddle--node case we will focus on
later and refer to {\cite{e3,e_snag,e_twist}} for this material but point out
that the simple pole singularity defined by $\check{\varphi} = \frac{1}{2 \pi
i}$ is a unit for this product and will be denoted by $\delta$ (in the Borel plane it is a unit for the convolution, thus acts as a Dirac at the origin for the Laplace transform of distributions or hyperfunctions).

\subsection{Alien operators; systems of alien derivations}

Let us consider an analytic function $\hat{\varphi}$ of the variable $\zeta$,
which can be analytically continued on a sector bisected by the positive real
axis, with isolated singularities at points $\omega_i$ on $\mathbbm{R}_{> 0}$
:
\[ 0 < \omega_1 < \omega_2 < \ldots < \omega_n < \ldots \]
The hypothesis on $\hat{\varphi}$ we just stated precisely means {\tmem{here}}
that $\hat{\varphi} (\zeta)$ can be analytically continued along any path
$\gamma$ starting from $0$ and avoiding any point in the set $\Omega = \{
\omega_1, \omega_2 \ldots \}$. We are going to define the alien operators for
such functions, referring for more general situations to the foundational
texts {\cite{e1,e3,e_snag}} and also to {\cite{sau_sn,sau_book}}.

For any $n \in \mathbbm{N}_{>}$, let us denote by $\hat{\varphi}_{\omega_n}^+$
the function which is the analytic continuation of the germ $\hat{\varphi}$ to
the point $\omega_n$, by following the positive real axis from below;
$\hat{\varphi}_{\omega_n}^+ (\xi)$ is in particular defined for $\xi$
belonging to a universal cover of a pointed disk $D_{\ast} \assign D
(\omega_n, \rho) - \{ \omega_n \}$centered at $\omega_n$.

We can define an operator $\Delta_{\omega_n}^+$ which ``measures'' the
singular behaviour of the function $\hat{\varphi}$ at $\omega_n$, by:
\[ \Delta_{\omega_n}^+ (\hat{\varphi}) (\zeta) \assign
   (\hat{\varphi}_{\omega_n}^+  (\zeta + \omega_n))^{\nabla} \]
where $\zeta$ is positive and close to $0$ (we thus extract the singularity of
$\hat{\varphi}_{\omega_n}^+ $ at the singular point $\omega_n$ after
translation of the variable in order to obtain a singularity {\tmem{at the
origin}}). 
For a resurgent function $\varphi^{\nabla}$ whose minor $\hat{\varphi}$ is regular at the origin, with isolated singularities at the $\omega_i$ on $\mathbbm{R}_{> 0}$, $\Delta_{\omega_n}^+$ is defined by the same formula.

These operators satisfy a simple commutation relation with the ordinary
derivation $\partial$, which in the Borel plane is the multiplication by $-
\zeta$ (of majors and minors), namely:
\[ [\Delta_{\omega}^+, \partial] = - \omega \Delta_{\omega}^+ \qquad (\forall
   \omega \in \Omega) \]
The $\Delta_{\omega_n}^+$ are not derivations of the algebra of resurgent
functions, but they satisfy the following property:
\[ \Delta_{\omega}^+  (\hat{\varphi} \ast \hat{\psi}) = \sum_{\omega' +
   \omega'' = \omega} \Delta_{\omega'}^+  (\hat{\varphi}) \Delta_{\omega''}^+ 
   (\hat{\psi}) \]
This corresponds to the fact that the $\Delta_{\omega_n}^+$ are the
homogeneous components of the Stokes automorphism, here in the direction $d
=\mathbbm{R}_{>}$, (with respect to the grading by the $e^{\omega_n z}$ for
alien operators acting on the relevant space of formal expansions: we are
sketchy here and refer to {\cite{sau_sn,sau_book}} for precise formulations)
when we deal with resurgent functions with exponential growth at $\infty$ .

The collection of the $\Delta_{\omega_n}^+$ can be used as building blocks to
construct other operators $\Gamma_{\omega_n}$ in these graded spaces of linear
endomorphisms, by formulas of the following type:
\[ \Gamma_{\omega_n} = \sum_{\omega_1 + \ldots + \omega_r = \omega_n}
   M^{(\omega_1, \ldots, \omega_r)} \Delta_{\omega_r}^+ \ldots
   \Delta_{\omega_1}^+ \]
Such a collection \ $M^{(\omega_1, \ldots, \omega_r)}$ is called a
{\tmem{mould}} and the collection of the $\Delta_{(\omega_1, \ldots,
\omega_r)}^+$, where \  $\Delta_{(\omega_1, \ldots, \omega_r)}^+ \assign
\Delta_{\omega_r}^+ \ldots \Delta_{\omega_1}^+$, a {\tmem{comould}}; basic facts about
moulds and comoulds are recalled below. 
When the coefficients satisfy suitable
symmetry properties, the $\Gamma_{\omega_n}$ will be derivations of the
algebras of resurgent functions on which they act: these are the {\tmem{alien
derivations}} and there are several systems of them, depending on the family
of coefficients $M^{(\omega_1, \ldots, \omega_r)}$.
When dealing with resurgent functions which have minors with isolated singularities $\omega$ over a given direction $d$ which is not necessarily $\mathbbm{R}_{> 0}$, alien operators ${\Delta_{\omega}}^+$ , $\Delta_{\omega}$ are defined by the same process (in case of minors which are not uniform at the origin, alien operators shall have to be indexed by points $\omega$ on $\mathbbm{C}_{\infty}$, we refer to \cite{e3,e_snag}).

The standard system of alien derivations $\Delta_{\omega}^{\tmop{stan}}$ are
defined by $M^{(\omega_1, \ldots, \omega_r)} \assign \frac{(- 1)^{r - 1}}{r}$
and are simply denoted by $\Delta_{\omega}$; they correspond to the
homogeneous components of the ``directional derivation'' which is the
logarithm of the Stokes diffeomorphism in the direction $d$ under
consideration -- which we suppose here to be the positive real axis.

There are many other systems of alien derivations, due to Ecalle
({\cite{e_twist}}), which are particularly relevant to the synthesis problem;
however, as we explain below, we have chosen to work in the present text with
the operators $\Delta_{\omega}^+$ and we refer to the appendix for brief
explanations on the other families of alien derivations, notably the
so--called organic ones $\Delta_{\omega}^{\tmop{org}}$ and to {\cite{e_twist}}
for detailed descriptions.

By pullback under the formal Borel transform, we get alien operators and in
particular derivations -- acting on algebras of formal series, denoted by the
same symbols $\Delta_{\omega}^+, \Delta_{\omega_{}}$.

Finally we can define, on sets of formal expansions involving at the same time
formal series in $z^{- 1}$ and exponential factors of the type $e^{\omega z}$
(so--called {\tmem{transseries}}, see e.g {\cite{e_twist,abs2019}}), the
``exponential bearing'' operators $ \dot{\Delta} _{\omega}^+ : = e^{-
\omega_{} z} \Delta_{\omega}^+$ and $ \dot{\Delta} _{\omega} : = e^{-
\omega_{} z} \Delta_{\omega}^{}$ which commute with the ordinary one
$\partial_{} = \frac{d}{d z}$.

\section{The resurgent study of the saddle--node}

\subsection{Analysis}

Saddle--nodes which have the simplest normal form correspond to non linear
differential equations in the resurgent variable $z$ of the following type
\[ (E) \qquad \partial_z y (z) = y (z) + b (z^{- 1}, y) \qquad (z \thicksim
   \infty, b \tmop{analytic}) \]
They have as formal normal form the elementary equation:
\[ (E^{\tmop{nor}}) \qquad \partial_z y (z) = y (z) \quad (z \thicksim \infty)
\]
The general solution of $E^{\tmop{nor}}$ being $y (z) = u e^z$ ($u \in
\mathbbm{C}$), when $\frac{1}{z} = x \thicksim 0$, the real phase space indeed
displays a saddle--type behaviour on the right and node--type one on the left.

\

Let $y (z, u) \in \mathbbm{C} [[z^{- 1}, u e^z]]$ be the formal integral --
namely the general formal solution of $(E)$; the resurgence properties of the
saddle--node are summed up in the following theorem:

\begin{theorem}
  (Ecalle {\cite{e3}}) The formal integral $y (z, u) = \sum_{n \geqslant 0}
  u^n e^{n z} y_n (z)$ is resurgent in $z$: the formal series $y_n$ are
  Gevrey--1, their Borel transforms have simple singularities (simple poles +
  logarithmic singularities) at the non zero integers and they have
  exponential growth in any non--real direction. We have moreover, for any
  given family of alien derivations {\tmstrong{the bridge equation}}:
  \[ \dot{\Delta}_n y (z, u) =\mathbbm{A}_n y (z, u) \quad n \in \Omega = \{ -
     1 \} \cup \mathbbm{N}^{\ast} \]
  Where $\mathbbm{A}_n = A_n u^{n + 1} \frac{\partial}{\partial u}$ ($A_n \in
  \mathbbm{C}$) is a derivation in the $u$ variable; the equality that casts a
  bridge between alien and ordinary calculus has to be understood
  component-wise:
  \[ \Delta_n (y_m) = (m - n) A_n y_{m - n} \quad \tmop{if} \quad 1 \leqslant
     n \leqslant m - 1 \quad (\tmop{and} 0 \tmop{else}) \]
  The family of numbers $(A_n)_{n \in \Omega}$ constitutes a complete set of
  analytic invariants: two equations that are analytically conjugate to the
  same normal form $E^{\tmop{nor}}$ have the same family of invariants and
  conversely, two equations which share the same family are analytically
  conjugate.
\end{theorem}

Once the resurgent character of the series involved is proved (beside e.g.
{\cite{e3}}, see also {\cite{sau_book}} and the extensive calculations in
{\cite{sau_sn}}), the bridge equation is easy to obtain with alien
{\tmem{derivations}} $\Delta_n$, by formal considerations involving the
Leibniz rule and homogenity properties satisfied by the
$\Delta_n$({\cite{e3,e_snag,sau_sn,sau_book}}).

Then, one can derive a version of the bridge equation for the action of the
operators $\Delta_n^+$: we get ordinary differential operators
$\mathbbm{A}_n^+$ in the $u$ variable which are no longer derivations, yet
each $\mathbbm{A}_n^+$ shares with the corresponding $\mathbbm{A}_n$ the same
property of homogeneity (of order $n$).\quad

Moreover, as the $\mathbbm{A}_n^+$ represent the homogeneous components of the
Stokes automorphism, {\tmem{their growth estimates in $n$, as linear
operators, are exponential}} ($\limsup \frac{\log (| A_n^+ |)}{n} < +
\infty$) where the $\mathbbm{A}_n$ associated to the standard alien
derivations generically display a superexponential growth
({\cite{e_snag,e_twist}}, see also the appendix). \

That property of the
$\mathbbm{A}_n^+$ make them very convenient to formulate the inverse problem.

\subsection{Synthesis}

Let us now tackle the inverse problem within the framework of resurgence, for
the saddle--node singularity -- and considering the simplest normal form, as
above. For this purpose, the data are a given family of ordinary differential
operators $(\mathbbm{A}_n^+)_{n \in \Omega = \{ - 1 \} \cup \mathbbm{N}_{>}}$,
which play the role of the ones to be obtained, through the bridge equation, by the
action of a family of alien operators $\Delta_n^+$ on the formal solution of
an analytic vector field that we wish to construct.

We will moreover suppose that $\mathbbm{R}_{>}$ is the only singular ray,
which amounts to supposing that $\mathbbm{A}_{- 1} = 0$ : this simplifying
assumption will enable us to focus on the main difficulties, but {\tmem{the
scheme works in the general case}} (see the last section).

Our aim will thus be to build an {\tmem{analytic}} vector field $X$ by a
conjugation to the normal form $X^{\tmop{nor}}$, through a change of
variables that can be chosen (\cite{martinet_ramis,sau_sn,sau_book}) as a fibered transformation
$\theta : (z, u) \longrightarrow (z, \varphi (z, u))$, which will
have the required moduli; the transformation $\theta$ will encode all the
information.

In fact, the main object which we shall construct will be an endomorphism
$\Theta$ of a space of formal series, which is equivalent to $\theta$; indeed,
there is a biunivoque correspondance between tangent to identity formal
diffeomorphisms $\theta$ and tangent to identity automorphisms $\Theta$ of
spaces of formal series ({\cite{e3,e_snag,e_twist,fms}}):
\[ \theta \mapsto \Theta \quad \tmop{with}\quad \Theta (f) \assign f \circ \theta
\]

All the following calculations will thus be performed at the operatorial level
and {\tmem{we are going to build}} $\Theta$ {\tmem{by taking as input a given
family}} $(\mathbbm{A}_n^+)_{n \geqslant 1}$ as above; for this, we write an
{\tmem{Ansatz}} and {\tmem{a priori}} express $\Theta$ as:
\[ \Theta = \sum_{\tmmathbf{n}^{} = (n_1, \ldots, n_r), n_i \in
   \mathbbm{N}_{>}} \mathcal{L}^{\tmmathbf{n}^{}} (z) \mathbbm{A}_{n_r}^+
   \ldots \mathbbm{A}_{n_1}^+ \]
\[ \  \]
The $\mathcal{L}^{\tmmathbf{n}^{}} (z)$ designate resurgent functions of the
variable $z$, indexed by sequences $\tmmathbf{n}^{} = (n_1, \ldots, n_r)$ of
nonnegative integers.

If the $\mathcal{L}^{\tmmathbf{n}^{}} (z)$ have nondecreasing valuations in
$n$, where $n = \| (n_1, \ldots, n_r) \| \assign n_1 + \ldots + n_r$, and
taking into account the homogeneity properties of the $\mathbbm{A}_{n_i}$, the
formal sum above will make sense in the space of operators acting on
$\mathbbm{C} [[z^{- 1}, u]]$ and indeed yield a tangent to identity linear
endomorphism. Now, such an object will solve our synthesis problem if the
following conditions are satisfied:
\begin{enumerate}
  \item The linear endomorphism $\Theta$ must be an automorphism of the
  {\tmem{algebra}} of formal series
  
  \item $X = \Theta X^{\tmop{nor}} \Theta^{- 1}$ must be analytic
  
  \item The moduli obtained by the action of the $\Delta_n^+$ on $X$ are the
  $\mathbbm{A}_n^+$ we started from
\end{enumerate}

The first point will follow from the symmetry properties of the mould
$\mathcal{L}^{\bullet}$: if it is {\tmem{symmetre}}l, then $\Theta$ will
indeed be a morphism of algebras, due to the {\tmem{cosymmetrelity}} of the
comould $\mathbbm{A}_{\bullet}^+$ (invertibility of $\Theta$ is immediate,
when $\mathcal{L}^{\varnothing} = 1$, as $\mathbbm{A}^+_{\varnothing} =
\tmop{Id}$).

Of course, if we had decided to take as input the $\mathbbm{A}_n$ associated
to some family of alien {\tmem{derivations}} $\Delta_n$ instead of the
$\mathbbm{A}_n^+$, the comould $\mathbbm{A}_{\bullet}$ would have been
{\tmem{cosymmetral}} and we would have required the {\tmem{symmetrality}} of
the mould to get the good algebraic properties we need.

The commutation relation $[\dot{\Delta} _n^+, \Theta] = - \Theta
\mathbbm{A}_n^+$, which expresses the bridge equation in operatorial form,
will be {\tmem{automatically satisfied}} if the endomorphism $\Theta$ is built
by using a mould $\mathcal{L}^{\bullet}$ with $\mathcal{L}^{\tmmathbf{\omega}}
= (- 1)^r \mathcal{G}^{\tmmathbf{\omega}} $, where $\mathcal{G}^{\bullet}$ is
a ``$\Delta$-friendly'' family of resurgent functions, in the following sense:
\[ \begin{array}{llll}
     \Delta_{n_0} \mathcal{G}^{n_1, \ldots, n_r} (z) & = & \mathcal{G}^{n_2,
     \ldots, n_r} (z) & (\tmop{if} n_0 = n_1)\\
     & = & 0 & (\tmop{if} n_0 \neq n_1)
   \end{array} \]

Moreover, all alien derivations will, by the very definition of $X$ act
trivially on it, which will entail the analyticity of $X$. But for this scheme
to work, we of course first have to {\tmstrong{prove that $\Theta$ defines an
endomorphism of the relevant space of resurgent functions -- and this is the
crux}}.

There are several technical difficulties to overcome, notably the following
one: as the data $\mathbbm{A}_{\tmmathbf{n}}^+$ involve composing ordinary
differential operators $\mathbbm{A}_{n_i}^+$, even if the family of operators
$\mathbbm{A}_{n_i}^+$ display an exponential growth, we will only get by brute
force estimates of the type $r! K^{n_1 + \ldots + n_r}$ for the compositions
$\mathbbm{A}_{n_r}^+ \ldots \mathbbm{A}_{n_1}^+$. The spurious $r!$ factor
will eventually prevent us to get normal convergence of the sums defining
$\theta$ in spaces of resurgent functions.

Thus, a clever reorganization of the expansion expressing $\Theta$ is
required and this will be provided by Ecalle's arbomould/coarmould formalism
({\cite{e_snag,fms,f_menous}}), explained in the section on moulds below. The
point is that we must ``reorder'' the mould--comould sums and consider
expansions over all forests $F$ decorated by nonnegative integers involving
armoulds $\mathcal{L}^{<} \assign (\mathcal{L}^F)$ and coarmoulds
$(\mathbbm{A}_F^+)$:
\[ \Theta = \sum_F \mathcal{L}^F \mathbbm{A}_F^+ \]
To sum up, {\tmstrong{the punchline amounts to finding an armould}}
$(\mathcal{L}^F)$ {\tmstrong{which is built out of ``$\Delta$-friendly''
resurgent monomials, with exponential bounds,}} for suitable semi--norms
defining the topology of the relevant topological vector space of resurgent
functions.

That requirement is highly non trivial and it was only in the late 90's that
such good families of (ar)moulds were found by Ecalle: the so--called
{\tmem{perilogarithms}}, among which the family of {\tmem{paralogarithms}}
stands out.

\subsection{Formal synthesis with hyperlogarithmic moulds}

The hyperlogarithmic mould $\mathcal{V}^{\bullet}$ satisfies exponential
growth estimates, in spaces of resurgent functions with tame behaviour at
infinity (see {\cite{e_snag}} and, for the case of the saddle--node, the
detailed calculations in {\cite{sau_sn}}).

These properties have in particular been used by Lappo-Danilevsky for his
beautiful partial solution of the Riemann--Hilbert problem with
hyperlogarithms (cf e.g. the Bourbaki lecture by Beauville {\cite{beau}}). It
is worth recalling here that Lappo-Danilevsky's scheme to give explicit
solutions to the inverse problem for linear systems involves a clever use of
the inverse function theorem, that works when the {\tmem{(monodromy) data are
close to the identity}}. Ecalle's scheme with paralogarithmic resurgence
monomials, explained below, applies to {\tmem{any (Stokes) data}}, thanks to the presence of a parameter $c$ which we can adjust.

The mould $\mathcal{V}^{\bullet}$, and some generalizations of it, is most
useful for the {\tmem{analysis}} of the resurgence properties of very general
dynamical systems and we refer to the foundational {\cite{e3,e_snag}} for
far-reaching results obtained with it for vector fields in any complex
dimension.

Already in the early 80's, J. Ecalle had introduced a ``$\Delta$--friendly''
companion mould $\mathcal{U}^{\bullet}$ (described below) to the mould
$\mathcal{V}^{\bullet}$ and implemented it to tackle the synthesis problem,
but he has shown that {\tmem{the objects thus synthesised with the help of
this hyperlogarithmic $\mathcal{U}^{\bullet}$ mould were in fact formal series
that are generically divergent}}.

We refer to {\cite{e_twist}} section 4 for a careful analysis of this
divergence and of the related specific resurgence properties, in particular
about the necessity to introduce some parameters in the process, because of
the ``multivaluedness'' of the relations between the ``invariants'' and the
``coinvariants'' (which are by definition the coefficients of the dynamical
systems synthesized, here by the use of hyperlogaritmic resurgence monomials).

\section{Elements of mould calculus}

\subsection{Moulds and comoulds}

\begin{definition}
  A mould $M^{\bullet}$ is a family of elements $M^{\tmmathbf{\omega}}$ of
  some commutative algebra $\mathcal{A}$ over $\mathbbm{C}$, indexed by the
  set $\Omega^{\ast}$ of sequences \ $\tmmathbf{\omega}= (\omega_1, \ldots,
  \omega_r)$ of elements of some set $\Omega$, also denoted as words $\omega_1
  \ldots \omega_r$ with letters in the aphabet $\Omega$.
\end{definition}

A mould can alternatively be seen as a linear function from the free vector
space spanned by sequences/words to the algebra $\mathcal{A}$; this point of
view can be quite fruitful (see e.g. \cite{efm,f_menous}), although we shall stick here to Ecalle's
definitions and notations.

In the applications, moulds are very often associated with elements
$B_{\tmmathbf{\omega}}$ of some bialgebra, to constitute formal sums; more
precisely, we shall deal with expressions of the following type:
\[ \Phi = \sum M^{\tmmathbf{\omega}} B_{\tmmathbf{\omega}} = \sum_{r \geqslant 0}
   \sum_{\tmmathbf{\omega}= \omega_1 \ldots \omega_r} \tmmathbf{}
   M^{\tmmathbf{\omega}} B_{\tmmathbf{\omega}} \]
The family of the $B_{\tmmathbf{\omega}}$ is called a comould and such an
expansion a mould--comould contraction; when a relevant condition for $\Omega$
of ``local finiteness'' is satisfied, such sums make sense on the completed
free associative algebra spanned by the $B_{\tmmathbf{\omega}}$, under an
hypothesis that the components of the mould and comould have reasonable
homogeneity properties.

If the $B_{\tmmathbf{\omega}}$ represent products of derivations of
$\mathcal{A}$ ($B_{\tmmathbf{\omega}} = B_{\omega_r} \ldots B_{\omega_1}$,
with the $B_{\omega_i}$ satisfying Leibniz identity), such an expansion $F$
will itself be an automorphism if the coefficients satisfy corresponding
symmetry properties, hence the following:

\begin{definition}
  A mould is called symmetral iff $M^{\varnothing} = 1$ and, for every pair of
  non--empty words $\tmmathbf{\omega}'$, $\tmmathbf{\omega}''$, we have
  \[ M^{\tmmathbf{\omega}'} M^{\tmmathbf{\omega}''} = \sum_{\tmmathbf{\omega}
     \in \tmop{sh} (\tmmathbf{\omega}', \tmmathbf{\omega}'')}
     M^{\tmmathbf{\omega}} \]
\end{definition}

Where $\tmop{sh} (\tmmathbf{\omega}', \tmmathbf{\omega}'')$ designates the set
of words $\tmmathbf{\omega}$ that can be obtained by shuffling
$\tmmathbf{\omega}'$ and $\tmmathbf{\omega}''$, counting multiplicities; thus,
for distinct elements $\alpha, \beta, \gamma$ we have
\[ M^{\alpha \beta} M^{\gamma} = M^{\alpha \beta \gamma} + M^{\alpha \gamma
   \beta} + M^{\gamma \alpha \beta} \]
and
\[ M^{\alpha \beta} M^{\beta} = 2 M^{\alpha \beta \beta} + M^{\beta \alpha
   \beta} \]
\begin{definition}
  A mould is called alternal iff $M^{\varnothing} = 0$ and, for every pair of
  non--empty words $\tmmathbf{\omega}'$, $\tmmathbf{\omega}''$, we have
  \[ 0 = \sum_{\tmmathbf{\omega} \in \tmop{sh} (\tmmathbf{\omega}',
     \tmmathbf{\omega}'')} M^{\tmmathbf{\omega}} \]
\end{definition}

As we shall deal with mould--comould contractions where the comould is made
with \ shall also need the following:

\begin{definition}
  A mould is called symmetrel iff $M^{\varnothing} = 1$ and, for every pair of
  non--empty words $\tmmathbf{\omega}'$, $\tmmathbf{\omega}''$, we have
  \[ M^{\tmmathbf{\omega}'} M^{\tmmathbf{\omega}''} = \sum_{\tmmathbf{\omega}
     \in \tmop{csh} (\tmmathbf{\omega}', \tmmathbf{\omega}'')}
     M^{\tmmathbf{\omega}} \]
\end{definition}

In this definition, ordinary shuffles are replaced by {\tmem{contracting
shuffles}} $\tmop{csh}$: we shuffle 2 words $\tmmathbf{\omega}'$,
$\tmmathbf{\omega}''$ and then also perform all the possible contractions
$\omega'_i, \omega''_{i + 1} \longrightarrow \omega'_i + \omega''_{i + 1}$ of
2 or more consecutive elements of $\tmmathbf{\omega}'$and
$\tmmathbf{\omega}''$ respectively; thus:
\[ M^{\alpha \beta} M^{\gamma} = M^{\alpha \beta \gamma} + M^{\alpha \gamma
   \beta} + M^{\gamma \alpha \beta^{}} + M^{\alpha (\beta + \gamma)} +
   M^{(\alpha + \gamma, \beta)} \]

\begin{definition}
  A mould is called alternel iff $M^{\varnothing} = 0$ and, for every pair of
  non--empty words $\tmmathbf{\omega}'$, $\tmmathbf{\omega}''$, we have
  \[ 0 = \sum_{\tmmathbf{\omega} \in \tmop{csh} (\tmmathbf{\omega}',
     \tmmathbf{\omega}'')} M^{\tmmathbf{\omega}} \]
\end{definition}

\

\begin{definition}
  A comould $B_{\bullet}$ is a family of differential operators
  $B_{\tmmathbf{\omega}}$ (or more generally of \ some cocommutative
  bialgebra) indexed by the set $\Omega^{\ast}$ of sequences \
  $\tmmathbf{\omega}= (\omega_1, \ldots, \omega_r)$ of elements of some set
  $\Omega$.
\end{definition}

Most of the times in the applications, the components $B_{\omega}$ will have a
certain degree of homogeneity $N_{\omega}$, as linear operators acting on
formal series (thus, in one dimension, $B_{\omega} (x^n) = \beta_{\omega} x^{n
+ N_{\omega}}$, with $\beta_{\omega} \in \mathbbm{C}$ and $N_{\omega} \in
\mathbbm{N}$).

\subsection{Operations}

There are 2 very natural operations, to be performed on mould/comould
contractions:
\begin{enumerate}
  \item We can change the family of coefficients, keeping the set of letters
  $B_{\tmmathbf{\omega}}$ fixed; this will give us the product $\times$ of
  moulds.
  
  \item We can change the set of letters $B_{\tmmathbf{\omega}}$, through the
  action of some mould $M^{\bullet}$, this way
  \[ B_{\bullet} \longrightarrow C_{\bullet}  \quad \nocomma \nocomma
     \tmop{with} C_{\omega_0} = \sum_{\| \tmmathbf{\omega} \| = \omega_0}
     M^{\tmmathbf{\omega}} B_{\tmmathbf{\omega}} \]
  and this will give the composition $\circ$ of moulds
  
  \ 
\end{enumerate}
\begin{definition}
  Let $M^{\bullet}$ and $N^{\bullet}$ be 2 moulds
  \begin{enumerate}
    \item The product $P^{\bullet} = M^{\bullet} \times N^{\bullet}$ is
    defined by
    \[ P^{(\omega_1, \ldots, \omega_r)} = \sum_{i = 0}^r M^{\omega_1 \ldots
       \omega_i} N^{\omega_{i + 1} \ldots \omega_r} \]
    \item The composition $Q^{\bullet} = M^{\bullet} \circ N^{\bullet}$ is
    defined by
    \[ Q^{(\omega_1, \ldots, \omega_r)} = \sum M^{(\| \tmmathbf{\omega}^1 \|,
       \ldots, \| \tmmathbf{\omega}^s \|)} N^{\tmmathbf{\omega}^1} \ldots
       N^{\tmmathbf{\omega}^s} \]
    Where the summation is over all partitions of $\tmmathbf{\omega}$ by
    non--empty words $\tmmathbf{\omega}=\tmmathbf{\omega}^1 \ldots
    \tmmathbf{\omega}^s$
  \end{enumerate}
\end{definition}

Both operations are associative but they are noncommutative as soon as
$\Omega$ has more than one element. The mould $1^{\bullet}$, which takes the
value $1$ for the empty sequence and $0$ else, is the inverse for the
multiplication; the mould $I^{\bullet}$ which takes the value $1$ for
sequences with a single letter and $0$ else is a composition inverse on the
set of moulds $M^{\bullet}$ with $M^{\varnothing} = 0$.

\subsection{Armoulds and coarmoulds}

The arborification/coarborification process appears in the context of
mould--comould contractions:
\[ \sum M^{\tmmathbf{\omega}} B_{\tmmathbf{\omega}} \]
When one composes ordinary differential operators in $\mathbbm{R}^N$or
$\mathbbm{C}^N$, trees naturally come in as combinatorial tools, as one can
already see in one dimension: let us consider functions $\varphi$ of one
complex variable $x$, denote by $\partial = \frac{d}{d x}$ and take a family
of operators $B_i$ which are derivations $B_i = b_i (x) \partial$; for any
$\mathcal{C}^{\infty}$ function $\varphi (x)$, the action of the composition
of $B_2$ and $B_1$ on $\varphi$ yields 2 terms:
\[ B_2 B_1 (\varphi) (x) = b_2 (x) (\partial b_1 (x)) \partial (\varphi) (x)
   + b_2 (x) b_1 (x) \partial^2 (\varphi) (x) \]
If we leave aside the observable $\varphi$ and work at the level of operators,
$B_2 B_1$ is the sum of one operator $B^{'}$ which is the output of the
``action of $B_2$ on $B_1$'', the resulting operator the acting on observables
and a second one $B^{''}$which corresponds to $B_2$ and $B_1$ each acting each
on the observables. Pictorially:
\[ \begin{array}{cc}
     B^{'} : \Sbunbdeux\quad & B^{''} : \Fbunbdeux
   \end{array} \]

\

When composing 3 operators, $P = B_3 B_2 B_1$ can be represented by the sum
of 6 ``forests'':
\[ \begin{array}{llllll}
    \Sbunbdeuxbtrois ;& B_1\quad \Sbdeuxbtrois ;& B_2\quad \ \Sbunbtrois;& B_3\quad \ \Sbunbdeux ;& \BbunFbdeuxbtrois ;& \Fbunbdeuxbtrois
   \end{array} \]

\

This observation was first made by Cayley in the middle of the
$19^{\tmop{th}}$ century ({\cite{cayley}}) and it has been used in a number of
works since then, and in particular been implemented in a Hopf--algebraic
formalism by Grossman and Larson (cf e.g. \cite{f_menous,fms,foissy_unter} for related combinatorial constructions).

Ecalle has developed a much richer and structured algebraic formalism by
considering operations and symmetries involved {\tmem{both}} for the families
of composition of differential operators and the families of coefficients with
which they are matched, in mould/comould expansions. We now state the
definitions and properties we shall need, referring for further information
and proofs to the foundational texts by Ecalle
{\cite{e_snag,e_twist,ecallevallet}} and also to
{\cite{f_menous,fms,menous_birkh,menous_q}} (in particular, all the required
terminology and facts regarding the combinatorics on trees and forest are
explained in detail in {\cite{fms}}).

\begin{definition}
  \
  
  An arborescent mould (armould, for short) $M^{<}$ is a family of elements
  of some commutative algebra $\mathcal{A}$, indexed by forests which are
  decorated by elements of a given set $\Omega$. The mould $M^{<}$ is called
  separative iff it is multiplicative on the monoid of forests:
  \[ M^{F' F''} = M^{F'} M^{F''} \hspace{3em} (\tmop{and} M^{\varnothing} =
     1_{\mathcal{A}}) \]
  \quad
  
  An arborescent comould (coarmould, for short) $B_{<}$ is a family of
  ordinary differential operators (or elements of some cocommutative
  bialgebra), indexed by forests which are decorated by elements of a given
  set $\Omega$. The comould $B_{<}$ is called coseparative iff it satisfies
  for every forest $F$:
  \[ B_F (f g) = \sum_{F = F' F''} B_{F'} (f) B_{F''} (g) \qquad (\tmop{and}
     B_{\varnothing} = \tmop{Id}) \]
\end{definition}

\

We now introduce the process of arborification/coarborification, which was
first presented out of the blue by Ecalle in {\cite{e_snag}}, through which we
reorder mould/comould contractions to express them as sums over decorated
forests -- thus called armould/coarmould contractions. For this purpose, we
shall need the following definition, to go from totally ordered sequences of
elements of a set $\Omega$ to $\Omega$--decorated forests.

\begin{definition}
  \
  
  For any forest $F$ decorated by elements $\omega_i$ and any totally ordered
  sequence $\tmmathbf{\omega}' = (\omega'_1 \ldots, \omega'_r)$, we denote \
  $\tmmathbf{\omega}' \preccurlyeq F$ iff there is a bijection from the set of
  summits of $F$ to the set of indices of {\tmstrong{$\tmmathbf{\omega}'$}},
  which is order preserving (wrt to the arborescent order of $F$ and the total
  order of the sequence, respectively) and transports the decorations:
  \[ \forall j \in \{ 1, \ldots, r \} \quad \omega'_j = \omega_i \quad
     \tmop{where} j = \sigma (i) \]
  For any forest $F$ decorated by elements $\omega_i$ and any totally ordered
  sequence $\tmmathbf{\omega}' = (\omega'_1 \ldots, \omega'_s)$, we denote \
  $\tmmathbf{\omega}' \precprec F$ iff there is a surjection $\pi$ from the
  set of summits of $F$ to the set of indices of
  {\tmstrong{$\tmmathbf{\omega}'$}}, which preserves the strict order and
  ``projects'' the decorations:
  \[ \forall j \in \{ 1, \ldots, s \} \quad \omega'_j = \sum_{\pi (i) = j}
     \omega_i \]
  
\end{definition}

\begin{definition}
  \
  
  The arborified of a given mould $M^{\bullet}$ is the armould $M^{<}$
  defined by
  \[ M^F \assign \sum_{\tmmathbf{\omega}', \tmop{such} \tmop{that}
     \tmmathbf{\omega}' \preccurlyeq F} M^{\tmmathbf{\omega}'} \]
  \quad The contracted arborified of a given mould $M^{\bullet}$ is the
  armould $M^{<}$ defined by
  \[ M^F \assign \sum_{\tmmathbf{\omega}', \tmop{such} \tmop{that}
     \tmmathbf{\omega}' \precprec F} M^{\tmmathbf{\omega}'} \]
\end{definition}

The arborified (resp. contracted arborified) of any symmetral (resp.
symmetrel) mould is \ separative ({\cite{e_snag,f_menous}}). Moreover, the
process of arborification is a morphism of algebras, the product for
arborescent moulds being the one introduced by Ecalle in 92 -- and
corresponding to the coproduct of Connes--Kreimer's Hopf algebra, when
armoulds are interpreted as characters ({\cite{f_menous, fms}}).

\begin{example}
  For $F =\BaFbc$ we have, for the arborified and the contracted arborified
  respectively:
  \[ M^F = M^{\Sabc} + M^{\Sacb} \]
  \[ M^F = M^{\Sabc} + M^{\Sacb}  + 2 M^{\Sabplusc} \]
\end{example}

\begin{definition}
  \
  
  A coarmould $B_{<}$ is called a coarborified of a given mould $B_{\bullet}$
  iff
  \[ B_{\tmmathbf{\omega}} = \sum_{\tmmathbf{\omega} \preccurlyeq F} B_F \]
  \quad A coarmould $B_{<}$ is called a contracted coarborified of a given
  mould $B_{\bullet}$ iff
  \[ B_{\tmmathbf{\omega}} = \sum_{\tmmathbf{\omega} \precprec F} B_F \]
  
\end{definition}

On the comould side, thus, as the process of coarborification involves a
decomposition, no uniqueness of the coarborified is to be expected.

There is however a recursive process, due to Ecalle, producing a coarmould
out of any given family of ordinary differential operators $(B_{\omega})$ with
reasonable homogeneity properties, yielding a coseparative comould $B_{<}$
whenever the inputs $B_{\omega}$ are all derivations {\tmem{or}} all satisfy
$B_{\tmmathbf{\omega}} (f g) = \sum_{\tmmathbf{\omega}=\tmmathbf{\omega}'
\tmmathbf{\omega}''} B_{\tmmathbf{\omega}'} (f) B_{\tmmathbf{\omega}''} (g)$;
this coarmould is such that in the former case $B_{<}$ is a coarborified of
the {\tmem{symmetral}} mould defined by $B_{(\omega_1, \ldots, \omega_r)}
\assign B_{\omega_r} \ldots B_{\omega_1}$ and in the latter a coarborified of
the {\tmem{symmetrel}} mould defined by the same formula.

The key property satisfied by the homogeneous coarborification is that the
coarmould it produces satisfies exponential growth estimates, as soon as its
building blocks $B_{\omega}$ do (e.g., in one dimension, $B_{\omega} (x^n) =
\beta_{\omega} x^{n + N_{\omega}}$, with $| \beta_{N_{\omega}} | \leqslant
H^{N_{\omega}}$, for some $H > 0$). This was stated, at the level of
operators, with very concise arguments to justify it, in the foundational
article {\cite{e_snag}} (see also section 11 of {\cite{e_twist}}). Later, a
proof of this crucial property was given, for the functions obtained by the
action of these operators on the coordinates, by F. Menous in
{\cite{menous_birkh, menous_q}} and implemented again in {\cite{f_menous,
fms}}; we refer to all these references for the precise definitions,
statements and proofs.

For any symmetral/cosymmetral (resp. symmetrel/cosymmetrel) mould contraction $\Phi = \sum M^{\tmmathbf{\omega}} B_{\tmmathbf{\omega}}$ , we have thus a systematic way to reorganize it as an armould/coarmould contraction, using the homogeneous coarborified and the arborified (resp. contracting arborified) of the involved mould by
  \[   \Phi = \sum_{F} M^F B_F
  \]

\section{Resurgence monomials}

\subsection{The problematics of resurgence monomials}

As the alien operators $\Delta_{\omega}^+$ and the alien derivations
$\Delta_{\omega}$ are powerful and efficient tools for characterizing the
singularities of the resurgent functions and exploring their behaviour on all
the sheets of their Riemann surfaces, it is most natural to search for
families of resurgent functions which are in some sense dual to these alien
operators.

More precisely, rather than looking for families ($M^{\omega} (z)_{\omega \in
\Omega}$, as the process of analytic continuation which underlies the very
definition of the alien operators entails non--commutativity, it is more
sensible to look for families $(M^{(\omega_1, \ldots, \omega_r)} (z)^{})$ of
resurgent functions indexed by sequences of elements of a given discrete set
$\Omega$ of complex numbers.

We thus wish to define moulds with values in spaces of resurgent functions,
which shall be as simple as possible and numerous enough to expand any given
resurgent function as an infinite sum of the $M^{(\omega_1, \ldots, \omega_r)}
(z)^{}$; moreover, we expect these moulds to have good symmetry properties --
symmetral or symmetrel, depending on the type of problems we wish to treat,
see {\cite{e_snag}}.

Beyond that, we need to have moulds that behave simply under the action of
the ordinary multiplication (which correspond to multiplication by $- \zeta$
in the Borel plane), and the alien operators. In fact, there is a tension
between these last 2 requirements: as explained in {\cite{e3}}, only trivial
functions will at the same time behave very simply with regard to both the
ordinary and the alien derivations; correspondingly, we shall have to relax
our exigencies and look for distinct families of resurgence monomials:
\begin{enumerate}
  \item Monomials with very simple properties wrt $\partial$, at the cost of
  satisfying a bit more complicated properties wrt the $\Delta_{\omega}$'s
  
  \item Monomials with very simple properties wrt the $\Delta_{\omega}$'s, at
  the cost of satisfying a bit more complicated properties wrt $\partial$
\end{enumerate}
Monomials of the first type, called $\partial$--friendly, are particularly
well suited for the {\tmem{analysis}} of the resurgence properties: namely
proving the resurgence properties and deriving the bridge equation. Those of
the second type, called $\Delta$--friendly, are the ones which will have to be
used for {\tmem{synthesis}} problems.

\

\subsection{Hyperlogarithms}

The most natural family of resurgence monomials is built
with {\tmem{hyperlogarithms}} ({\cite{e1,e3,e_snag,sau_sn}}). We define a
mould $\mathcal{V}^{\bullet}$, by recurrence on the length of the sequence
(with $\mathcal{V}^{\varnothing} = 1^{}$), by the following formula :

\[ (\partial_z + \| \tmmathbf{\omega} \|) \mathcal{V}^{\tmmathbf{\omega}} (z)
   = - z^{- 1} \mathcal{V}^{\omega_1, \ldots, \omega_{r - 1}} (z) \]
(we recall that $\| (\omega_1, \ldots, \omega_r) \| \assign \omega_1 + \ldots
+ \omega_r$)

In the Borel plane, this relation becomes (for minors):
\[ (- \zeta + \| \tmmathbf{\omega} \|) \hat{\mathcal{V}}^{(\omega_1, \ldots,
   \omega_r)} (\zeta) = - \int_0^{\zeta} \hat{\mathcal{V}}^{(\omega_1, \ldots,
   \omega_{r - 1})} (s) d s \]
The mould $\mathcal{V}^{\bullet}$ takes its values in the space of so--called
simple resurgent functions: at each of its singular point (for any given
sequence, there is a finite number of them, by construction), the singularity
is logarithmic plus a simple pole; moreover, $\mathcal{V}^{\bullet}$ satisfies
the shuffle relations:
\[ \mathcal{V}^{\tmmathbf{\omega}'} (z)^{} \mathcal{V}^{\tmmathbf{\omega}''}
   (z)^{} = \sum_{\tmmathbf{\omega} \in \tmop{sh} (\tmmathbf{\omega}',
   \tmmathbf{\omega}'')} \mathcal{V}^{\tmmathbf{\omega}} (z) \]

By their very definition, the $\mathcal{V}^{\tmmathbf{\omega}} (z)$ are
``$\partial$--friendly'': their behaviour under the ordinary derivation
$\partial$ is most simple. With respect to the action of alien derivations, it
is more involved and we have the following formulas for the action of the
standard alien dervivations $\Delta_{\omega}$, implying complex coefficients
$V^{\tmmathbf{\omega}}$ ({\cite{e1,e_snag,e_twist}}):
\[ \Delta_{\omega_0} \mathcal{V}^{\tmmathbf{\omega}} (z) = \sum_{\omega_1 +
   \ldots + \omega_i = \omega_0} V^{\omega_1, \ldots, \omega_i}
   \mathcal{V}^{\omega_{i + 1}, \ldots, \omega_r} (z) \]

\[ \Delta_{\omega_0} \mathcal{V}^{\tmmathbf{\omega}} (z) = 0 \qquad \tmop{if}
   \omega_1 + \ldots + \omega_i \neq \omega_0, \forall i \in \{ 1, \ldots, r
   \} \]

\

We then make use of the family of coefficients $V^{\tmmathbf{\omega}}$ (which
are built with multiple zeta values, see {\cite{e1}}, {\cite{sau_sn}}) to
construct the ``hyperlogarithmic $\Delta$--friendly mould $\mathcal{U}$''.

If we collect the $V^{\tmmathbf{\omega}}$'s for all words
$\tmmathbf{\omega}$, we indeed get a mould, which is $\mathbbm{C}$--valued;
this mould is alternal, it has an inverse $U^{\bullet}$ {\tmem{for the
operation of composition}} $\circ$ and we then define $\mathcal{U}$ by:
\[ \mathcal{U}^{\bullet} =\mathcal{V} \circ U^{\bullet} \]
The mould $\mathcal{U}^{\bullet}$ (beside {\cite{e1}}, detailed calculations
can be found in {\cite{sau_sn}}) is a symmetral mould with values in the space
of simple resurgent function and it satisfies:

\[ \begin{array}{llll}
     \Delta_{\omega_0} \mathcal{U}^{\omega_1, \ldots, \omega_r} (z) & = &
     \mathcal{U}^{\omega_2, \ldots, \omega_r} (z) & (\tmop{if} \omega_0 =
     \omega_1)\\
     & = & 0 & (\tmop{if} \omega_0 \neq \omega_1)
   \end{array} \]
For short, these relations expressing the behaviour of the
$\mathcal{U}^{\omega_1, \ldots, \omega_r}$ under the action of the alien
derivations, and similar ones for other families of operators, will be called
``orthogonality relations''. It is important to keep in mind that they depend
upon the particular set of alien operators we use to express them: here, it is
the family of standard alien derivations $\Delta_{\omega}$ but we can also use
other families: e.g. the organic alien derivations
$\Delta_{\omega}^{\tmop{org}}$, or the $\Delta_{\omega}^+$ and in this last
case, the corresponding monomials will be denoted by ${}^+ \mathcal{U}$.

\

With regards to the ordinary derivation $\partial$, the monomials
$\mathcal{U}$ satisfy the following resurgence relations:
\[ (\partial_z + \| \tmmathbf{\omega} \|) \mathcal{U}^{\tmmathbf{\omega}} (z)
   = \sum_{\omega_1 + \ldots + \omega_i = \omega_0} \mathcal{U}^{\omega_1,
   \ldots, \omega_i}  (z) U^{\omega_{i + 1}, \ldots, \omega_r} z^{- 1} \]
where the $U^{\tmmathbf{\omega}}$'s are complex constants.

The $\Delta$-- friendly mould $\mathcal{U}$ will for short be called
hyperlogarithmic, as it stems from the hyperlogarithmic $\partial$--friendly
mould $\mathcal{V}$.

\begin{remark}
  The reason for the divergence of the objects which are synthesized with the
  use of the ``hyperlogarithmic $\mathcal{U}$'' can in fact be traced to the
  operation of taking an inverse for the composition of moulds, in the process
  of the construction of the mould $\mathcal{U}$ from the mould $\mathcal{V}$.
  Indeed, both the product $\noplus \times$ and the composition $\circ$
  respect the property of exponential growth of moulds and so does the
  operation of inversion with respect to the product, but the calculation of
  the composition inverse is a very costly operation (already at the level of
  ordinary moulds, and also for arborescent ones) and that very fact
  eventually entails divergence of the constructions involving the
  hyperlogarithmic $\mathcal{U}$ mould for the synthesis problem. 
\end{remark}

\section{Paralogarithmic resurgent monomials and synthesis}

\subsection{Paralogarithms}

We shall now present, following Ecalle, a family of resurgence monomials that
possess all the required properties, notably the crucial one regarding
exponential growth. These monomials are built from the hyperlogarithmic ones;
they require a parameter -- but as we have mentioned above, some parameters
shall be expected in this process -- and the striking fact is that one is
enough and that this family (conveniently generalized, as explained in the
next section) will be truly universal : it will enable to solve all inverse
problems, for dynamical systems displaying resurgence.

We have described in the previous section the ``standard procedure'' to get
the $\Delta$--friendly family of monomials $\mathcal{U}$, from the
hyperlogarithmic one $\mathcal{V}$. In fact, there is a compact formula,
already given in {\cite{e1}}, that expresses the monomials ${}^+
\mathcal{U}^{}$ as a multiple integral:
\[ {}^+ \mathcal{U}^{\omega_1 \ldots \omega_r} (z) = \int_0^{\infty} \ldots
   \int_0^{\infty} \frac{\exp (- \omega_1 y_1 - \ldots - \omega_r y_r)}{(y_r -
   y_{r - 1}) \ldots (y_2 - y_1) (y_1 - z)} d y_1 \ldots d y_r \]
The integration is performed bypassing every $y_i$ from under
$\mathbbm{R}_{>}$ (or rather performing all integrations on distinct
directions $\theta_i$ that are slightly under the positive real axis); we have
of course a corresponding formula for the monomials ${}^-
\mathcal{U}^{\omega_1 \ldots \omega_r} (z)$). This formula is valid in any of
the 3 models of the resurgent functions involved, but the term $\frac{1}{y_1 -
z}$ has to be understood accordingly ({\cite{e_twist}}, section 10):
\begin{enumerate}
  \item in the formal plane as the series $\sum_{n = 0}^{\infty} z^{- n - 1}
  y_1^n$
  
  \item in the geometric models as a function of the complex variable $z$
  
  \item in the Borel plane as the function $\exp (- y_1 \zeta)$
\end{enumerate}
\begin{remark}
  The geometric models depend by definition upon the choice of a direction $d$
  bisecting a sector of opening $\pi$ on which the given function is {\tmem{a
  priori}} defined. In particular, if we express ${}^+ \mathcal{U}^{\omega_1
  \ldots \omega_r} (z)$ in a sector bisected by a direction $d_+$ (resp.
  $d_-$) slightly under (resp. above) the positive real axis, we shall keep a
  tag for this; we can keep the notation adopted in {\cite{e_twist}}, section
  10 and denote these sectorial incarnations of \ ${}^+ \mathcal{U}^{\omega_1
  \ldots \omega_r}$ \ by ${}^{\eta, +} \mathcal{U}^{\omega_1 \ldots \omega_r}
  (z)$, where $\eta = +$ or $-$ respectively.
\end{remark}

To introduce the paralogarithmic resurgence monomials, we reconsider now, here
for points $\omega_i$ on the positive real axis, the integral formulas
expressing the (exponential--bearing) hyperlogarithmic resurgent monomials
${}^+ \mathcal{U}e$, written in the following way:
\newpage

\[ 
     {}^+ \mathcal{U}e^{\omega_1 \ldots \omega_r} (z) =
 \]
 \[
      \int_0^{\infty} \ldots \int_0^{\infty} \frac{\exp (- \omega_1 y_1)
     \ldots \exp (- \omega_r y_r)}{\exp (- \omega_1 z) \ldots \exp (- \omega_r
     z)} \frac{1}{(y_r - y_{r - 1}) \ldots (y_2 - y_1) (y_1 - z)} d y_1 \ldots
     d y_r
 \]

The idea is to start from this formula and {\tmem{modify the exponential
integration kernel}} in such a way as to preserve the asymptotics and the
symmetry properties, in order to ensure that the new functions we build shall
indeed be resurgent and orthogonal to the corresponding alien operators but
also that they possess the crucial property of exponential growth.

For that purpose, we introduce now a {\tmem{nonnegative real parameter $c$}},
and consider the following function:
\[ g_{c, \omega} (y) \assign \exp (- \omega y - \omega c^2 y^{- 1}) \]
Along with $g_{c, \omega} (y)$, we shall also have to deal with the function
$f_{c, \omega}$ which is its Laplace transform:
\[ \begin{array}{lll}
     f_{c, \omega} (x) & = & \int_0^{+ \infty} e^{- x y} g_{c, \omega} (y) d y
   \end{array} \]
Accordingly, $g_{c, \omega}$ is the Borel transform of $f_{c, \omega}$:
\[ g_{c, \omega} (y) = \frac{1}{2 \pi i}  \int_{- i \infty}^{+ i \infty} e^{+
   x y} f_{c, \omega} (x) d x \]

\begin{definition}
  The paralogarithmic resurgent monomials are defined by:
  \[ 
       {}^+ \mathcal{U}_{} e_c^{\omega_1 \ldots \omega_r} (z) :=
   \]
 \[ \int_0^{\infty} \ldots \int_0^{\infty} \frac{g_{c, \omega_1} (y_1)
       \ldots g_{c, \omega_r} (y_r)}{g_{c, \omega_1} (z) \ldots g_{c,
       \omega_{_r}} (z)} \frac{1}{(y_r - y_{r - 1}) \ldots (y_2 - y_1) (y_1 -
       z)} d y_1 \ldots d y_r
 \]
\end{definition}

Alongside the exponential--bearing ${}^+ \mathcal{U}e_c^{\omega_1 \ldots
\omega_r} (z)$ we shall of course have to consider the monomials ${}^+
\mathcal{U}_c^{\omega_1 \ldots \omega_r} (z)$, which satisfy:
\[ \mathcal{U}_c e^{\omega_1 \ldots \omega_r} (z) = {}^+
   \mathcal{U}_c^{\omega_1 \ldots \omega_r} (z) e^{\| \omega_1 \ldots \omega_r
   \| z} \]
In fact, for the proofs of the main properties, it will be more convenient to
work with the auxiliary functions ${}^+ \mathcal{U}a_c^{\omega_1 \ldots
\omega_r} (z)$ (introduced in {\cite{e_twist}}), related to
$\mathcal{U}e_c^{\omega_1 \ldots \omega_r} (z)$ in the following way:
\[ \mathcal{U}e_c^{\omega_1 \ldots \omega_r} (z) = {}^+
   \mathcal{U}a_c^{\omega_1 \ldots \omega_r} (z) e^{\| \omega_1 \ldots
   \omega_r \| z + c^2 \| \omega_1 \ldots \omega_r \| z^{_{- 1}}} \]
Our working formula, with the same integration rule, will thus be:
\[ {}^+ \mathcal{U}a_c^{\omega_1 \ldots \omega_r} (z) = \int_0^{\infty} \ldots
   \int_0^{\infty} \frac{g_{c, \omega_1} (y_1) \ldots g_{c, \omega_r}
   (y_r)}{(y_r - y_{r - 1}) \ldots (y_2 - y_1) (y_1 - z)} d y_1 \ldots d y_r
\]

\

\begin{remark}
  For $c = 0$ , of course, the defining formula of the $\mathcal{U}_c (z)$
  boils down to hyperlogarithms but the asymptotics properties completely
  change as soon as $c > 0$ , {\tmem{}}which we will implicitely assume in
  what follows, when working with the $\mathcal{U}_c (z)$ monomials.
\end{remark}

\subsection{Schemes for the proofs}
As was pointed out in the introduction, the schemes for the proofs of the properties of paralogarithmic resurgence monomials are due to Ecalle and they are clearly indicated in the article \cite{e_twist}. That reference, however, contains a wealth of material on the topic of synthesis but also on other related important topics, which makes it not easily accessible to the readers who have no prior familiarity with mould calculus (this may explain why these beautiful objects had not yet been digested by the dynamical systems community, or beyond); we urge the reader to focus on sections $4,5,10,11$ of that reference (cf in particular subsection 11.3).

To work on the proofs, we have the choice between 2
expressions: the defining one with integration over the $y$ variables or
another formula, deduced from it through the Borel transform, namely :
\[
 \mathcal{U}a_c^{\omega_1 \ldots \omega_r} (z) =
 \] 
\[
     \int_{- i \infty}^{+ i \infty} \ldots \int_{- i \infty}^{+ i \infty}
     \sigma_+ (\widehat{x_1}) \sigma_- (\widehat{x_2}) \ldots \sigma_-
     (\widehat{x_r}) f_{c, \omega_1} (x_1) \ldots f_{c, \omega_r} (x_r) e^{\|
     \tmmathbf{\omega} \| z}_{} d x_1 \ldots d x_r
\]

\

In this expression, $\sigma$ designates Heaviside's step function
$1_{\mathbbm{R}_+}$ and, for any sequence $x_1, \ldots, x_r$ and index $i \in
\{ 1, \ldots, r \}$, $\hat{x}_i \assign \sum_{i \leqslant j \leqslant r} x_j$

For the proofs of the required properties of the monomials, we can take
advantage of the 2 multiple integrals; in particular, as explained in section
10 of {\cite{e_twist}}, both expressions can be used to obtain the crucial
exponential growth of the arborified moulds.

\

1) The resurgent character of the $\mathcal{U}_c^{\omega_1 \ldots \omega_r}
(z)$ is checked using the ``$y$-integrals'', by recurrence on $r$.

\

2) The orthogonality of the ${}^+ \mathcal{U}_c^{\omega_1 \ldots \omega_r}
(z)$ \ to the $\Delta_{\omega}^+$ is a direct consequence of the fact that the
${}^+ \mathcal{U}_c^{\omega_1 \ldots \omega_r} (z)$ have the same asymptotics
than the hyperlogarithmic ${}^+ \mathcal{U}^{\omega_1 \ldots \omega_r} (z)$,
because for any $c$, $\log (g_{c, \omega} (y)) \sim - y, \tmop{when} y \sim +
\infty \tmop{on} \mathbbm{R}_{>}$

\

3) The symmetrelity of the mould ${}^+ \mathcal{U}_c^{\bullet} (z)$ can be
checked on the $y$--integrals

\

4) To prove the exponential growth property, we can use the $x$--integrals.
For every positive $\omega$ we have:
\[ g_{c, \omega} (y) \leqslant \exp (- 2 \omega c) \qquad (\forall y > 0) \]
and for complex values of $y$ we have a saddle--point at $y = c$ (see also the estimates for the integrals in section 4 of \cite{cv} and thoses in appendix C of \cite{gmn}). The lack of
such a property for $c = 0$ of course corresponds to the inadequacy of the
hyperlogarithmic $\mathcal{U}$ for this scheme to yield convergent series as
output.

As the product $\sigma_+ (\widehat{x_1}) \sigma_- (\widehat{x_2}) \ldots
\sigma_- (\widehat{x_r})$ is obviously always $0$ or $1$, we indeed get an
exponential growth estimate:
\[ | \mathcal{U}a_c^{\omega_1 \ldots \omega_r} (z) | \leqslant K^{\omega_1 +
   \ldots + \omega_r} \]
which is valid uniformly for any $z$ in a sector of opening $< \pi$ , centered
on any direction except $\mathbbm{R}_{>}$: we indeed work {\tmem{in geometric
models}}, perform Laplace transforms on directions $d \neq \mathbbm{R}_{>}$
and use the saddle--point method to obtain the inequality -- with a constant
$K$ explicitely depending on the parameter $c$ in the following way:
\[ K \leqslant K_0 e^{- \alpha c} \qquad \]
with $K_0$ independent of $c$, and $\alpha > 0$.

Now, as we have seen, the ultimate estimate that we shall need is one of the
same type for the contracted arborified of the mould ${}^+ \mathcal{U}a$ (the
ones for ${}^+ \mathcal{U}$ will follow easily: they differ by a factor which
is an {\tmem{exponential in}} $z^{- 1}$, which corresponds in the Borel plane
to a convolution by a constant of resurgence).

This is the crucial point and it is settled by the fact that we have for the
(contracted) arborified $\mathcal{U}a^{<}$ an integral formula of exactly the
same shape, but the sums $\sum_{i \leqslant j \leqslant r} x_j$ are replaced
by sums $\sum_{i \preccurlyeq j} x_j$ where $\preccurlyeq$ \ designates the
arborescent order, for any given forest decorated by the $x_i$.

{\tmstrong{This point is essential}} : without such an lemma, there is no
chance of getting any estimate better than $r! C^{x_1 + \ldots + x_r}$ for a
given forest.

Here lies the true ``miracle'' of the arborification technique: it yields
closed formulas for the arborified of ``moulds of practical interest''
({\cite{e_snag,e_twist}}) and it is thanks to these closed formulas that the
required exponential growth can be checked. We thus give the crucial:

\begin{lemma}
  For any forest $F$, with decorations $x_i$, we have for $\mathcal{U}a_c^F
  (z)$ the following expression:
  \[ \int_{- i \infty}^{+ i \infty} \ldots \int_{- i \infty}^{+ i \infty}
     \sigma_+ (\widehat{x_1}) \sigma_- (\widehat{x_2}) \ldots \sigma_-
     (\widehat{x_r}) f_{c, \omega_1} (x_1) \ldots f_{c, \omega_r} (x_r) e^{\|
     \tmmathbf{\omega} \| z}_{} d x_1 \ldots d x_r \]
\end{lemma}

where for any $i$, $\hat{x}_i$:= \ $\sum_{i \preccurlyeq j} x_j$, with
$\preccurlyeq$ designating the arborescent order

\

Of course, with this lemma, the exponential growth is obtained for
$\mathcal{U}a_c^F (z)$ as for $\mathcal{U}a_c^{\bullet} (z)$.

\

\subsection{The paralogarithmic synthesis of analytic saddle--nodes}

We collect now all the tools at our disposal to implement the synthesis of
saddle--nodes with paralogarithmic resurgence monomials. For the sake of
simplicity, we still \ consider families of given invariants with
$\mathbbm{R}_{>}$ as only singular direction in the Borel plane but we stress
that the method works in the general case.

We thus take as data any family $\left( \mathbbm{A}_n^+ = A_n u^{n + 1}
\frac{d}{d u} \right)_{n \in \mathbbm{N}_{>}}$ of exponential growth, which we
view as a family of invariants obtained through the bridge equation, by using
the operators $\Delta_n^+$. We consider the comould $\mathbbm{A}^{+ \bullet}$:
\[ \mathbbm{A}_{n_1, \ldots, n_r}^+ =\mathbbm{A}_{n_r}^+ \ldots
   \mathbbm{A}_{n_1}^+ \]
Plugging in the paralogarithmic monomials ${}^+ \mathcal{U}e_c$, we get a
formal operator
\[ \Theta_c \assign \sum_{(n_1, \ldots, n_r)} (- 1)^r \mathcal{U}e_c^{n_1,
   \ldots, n_r} \mathbbm{A}_{n_1, \ldots, n_r}^+ \]
As the mould $\mathcal{U}e_c^{\bullet}$ is symmetrel, the operator $\Theta_c$
is an automorphism of the algebra of formal series in $(z, u)$ that yields,
when applied to the variables $z, u$ themselves, a fibered formal
normalization transformation:
\[ \theta_c \assign (z, u) \longrightarrow (z, \varphi_c (z, u)) \]
The operator $\Theta_c$ can be re--expressed as a sum over all forests $F$
decorated by positive integers
\[ \Theta_c \assign \sum_F (- 1)^r  {}^+ \mathcal{U}e_c^F \mathbbm{A}_F^+ \]
Finally, we consider the derivation $X_c$, acting {\tmem{a priori}} on spaces
of formal series, defined by:
\[ X_c = \Theta_c X^{\tmop{nor}} \Theta_c^{- 1} \]
By what we have just seen, $X_c$ defines an operator that acts on spaces of
holomorphic functions in the $z$ variable, on sectors at $\infty$, with
uniform bounds: for any ray $d$ except the positive real axis, $X_c$ induces
an endomorphism of the space of holomorphic functions, on this sector,
{\tmem{because we have exponential growth estimates both for the armould}}
$({}^+ \mathcal{U}e_c^F)$ {\tmem{and the coarmould}} $(\mathbbm{A}_F^+)$,
hence normal convergence of the sums expressing $\Theta_c$ in the relevant
spaces of holomorphic functions in sectors, with uniform exponential bounds at
$\infty$.

Moreover, the Stokes phenomenon for $X_c$ in the critical direction
$\mathbbm{R}_{>}$ is trivial because by its very construction all the alien
derivatives of the components of $X_c$ at the positive integers vanish (at the
operatorial level, we have $[X_c, \Delta_n] = 0, \forall n \in
\mathbbm{N}_{>}$)! We can now conclude that $X_c$ is analytic.

On the other hand, the commutation relation which we just used to define
$X_c$ is no other than the expression of the bridge equation at the level of
operators; thus $X_c$ has the prescribed family of analytic invariants
$(\mathbbm{A}_n^+)_{n \in \mathbbm{N}_{>}}$.

We are done: when expressed in the original variables $x, y$ near the origin,
$X_c$ is a vector field with analytic coefficients; we have achieved the
synthesis of an analytic vector field with saddle--node singularity that has
the prescribed set of invariants.

\begin{remark}
  We insist that the scheme that we have just described for the saddle--node
  case will work in full generality for the effective synthesis of any
  dynamical system displaying resurgence features ({\cite{e_twist}}, sections
  4, 10 and 11). Having at our disposal a family of good resurgence
  monomials, it can, in the same way, be plugged in mould--comould expansions
  to treat any inverse problem, hence this synthesis process can be called
  canonical -- see the last section. 
  We refer to the examples given in \cite{e_twist}, for various classes of dynamical systems with discrete or continuous time (see also
  {\cite{menous_q}} for normalization results of some non--linear
  $q$--difference equations within the (ar)mould formalism, the synthesis scheme also has a vocation to apply to these): all what is
  needed is a precise resurgent analysis yielding some form of the bridge
  equation, the shape of which will depend upon the class of functional
  equation under study.
\end{remark}

\begin{remark}
  Writing for example $X_c = \Theta_c (X^{\tmop{nor}} \Theta_c^{- 1})$ the
  {\tmem{analytic}} object $X_c$ thus appears as the product of 2
  {\tmem{divergent--resurgent}} objects, in the space of linear endomorphisms
  acting on formal series in 2 variables.
\end{remark}

\section{Other examples and outlook}

\subsection{More general families of paralogarithmic monomials}

We have presented in the previous section paralogarithmic resurgence monomials
$(\mathcal{U}_c^{\omega_1 \ldots \omega_r})$ indexed by sequences of positive
integers; to deal with more general indexing sets $\Omega$, we shall need the
following:

\begin{definition}
  For any complex number $\omega$, and any positive parameter $c$ we define
  \[ g_{c, \omega} (y) \assign \exp (\omega y - c^2 \bar{\omega} y^{- 1}) \]
\end{definition}

With this enhanced definition ({\cite{e_twist}}), these
$\mathcal{U}_c^{\omega_1 \ldots \omega_r} (z)$ will enable us to deal with
synthesis problems involving any discrete set $\Omega$ corresponding to
singularities in the Borel plane, by following the same scheme, for dynamical
systems that have the simplest normal form ({\cite{e_twist}}).

The paralogarithmic resurgence monomials $(\mathcal{U}_c^{\omega_1 \ldots
\omega_r})$ moreover satisfy 2 important properties:

\begin{proposition}
  {\tmdummy}
  
  \begin{enumerate}
    \item The $\mathcal{U}_c^{\omega_1 \ldots \omega_r} (z)$ induce real
    functions on the real line
    
    \item The $\mathcal{U}_c^{\omega_1 \ldots \omega_r} (z)$ obey the
    following homogeneity relations, in the multiplicative models (geometric
    or formal), they are invariant under dilations:
    \[ (y, c, \omega_i) \longrightarrow (l y, l c, l^{- 1} \omega_i) \qquad
       \forall l > 0 \]
    \[ \  \]
  \end{enumerate}
\end{proposition}

The first point of the proposition is an immediate consequence of the
definition of the $\mathcal{U}_c^{\omega_1 \ldots \omega_r} (z)$; however
elementary, it is quite important if we want to {\tmem{synthesize dynamical
systems with real coefficients}}, when we plug the monomials in expansions
involving $\mathbbm{A}_{\tmmathbf{\omega}}^+$ with real coefficients.

The second point is easy to check and is notably relevant when one wants to
explore the dependance in $c$ of the dynamical systems synthesized with these
monomials.

\

In order to deal with more general normal forms which involve some ramified
powers $z^{\sigma}$, we need to introduce ramified paralogarithms built with
the functions:
\[ g_{c, \omega, \sigma} (y) \assign \exp (\omega y - c^2 \bar{\omega} y^{- 1}
   + \sigma \log (y)) \]

Finally, in order to synthesize general dynamical systems involving several
critical times, we we must have recourse to ``polycritical ramified
paralogarithmic monomials''; these objects are also defined in
{\cite{e_twist}}, they are outside the scope of the present text.

\

\begin{remark}
  The simplifying assumption we had made (namely $A_{- 1} = 0$) in presenting
  for the simplest normal form of saddle--nodes the general scheme of
  synthesis with paralogarithmic resurgence monomials had the effect that any
  $c > 0$ is sufficient to ensure convergence of the armoulds/coarmould
  expansions; without this assumption, the scheme works and yields analytic
  objects when the parameter $c$ is large enough (see {\cite{e_twist}},
  sections 10-11).
\end{remark}

\subsection{The linear case}

Resurgent functions, alien and mould calculus were introduced and developped
to solve problems of moduli for non--linear dynamical systems, yet these
concepts can prove quite useful already in linear problems -- with less
technical complications. For the synthesis problem notably, paralogarithmic
resurgence monomials entail a systematic and effective procedure to build
analytic linear dynamical systems with arbitrary Stokes data: they are thus
relevant to tackle the Riemann--Hilbert problem.

If we consider an analytic or meromorphic linear dynamical system $X$ in any
complex dimension $\nu$, with a normal form $X = \sum_{i = 1}^{\nu} \lambda_i
x_i \partial_{x_i}$ (for simplicity, we restrain to systems with trivial
formal monodromy), the fundamental matrix solution of $X$ involves Gevrey--1
and generically divergent components, which are resurgent with singularities
in the Borel plane at points $\omega_{i j}$ := $\lambda_i - \lambda_j$, $1
\leqslant i, j \leqslant \nu$, $i \neq j$ ({\cite{e3,e_twist}}).

As we are dealing with linear equations -- and thus a finite number of
singularities, growth conditions for the coefficients produced by the bridge
equation is of course not a concern and we can take as input for the solution
with resurgence monomials of the synthesis problem, any family $(A_{\omega_{i
j}})$ corresponding to the action of the $\Delta_{\omega_{i j}}^+$ or whatever
system of alien derivations $\Delta_{\omega_{i j}}$ -- in particular the
standard ones.

The process yields an analytic linear dynamical system $X_c$ with prescribed
bridge equation (or equivalently Stokes) data, depending on a parameter $c$,
which must be chosen large enough, depending on the values of the data
$(A_{\omega_{i j}})$ for the mould expansions to converge in the relevant
spaces of analytic functions.

The arborification/coarborification process is not needed for linear systems:
the mould/comould expansions {\tmem{with paralogarithmic resurgence
monomials}} already converge (for suitably large $c$), but the corresponding
expansions with hyperlogarithmic resurgence monomials diverge, except when the
data $(A_{\omega_{i j}})$ are small, or in otherwise particular situations.

The dependence on the parameter $c$ can be exploited, with the help of the
many formulas given in section 12 of {\cite{e_twist}}, to tackle questions of
deformations for the linear dynamic systems thus synthesized.

\begin{remark}
  It is quite striking that, in order to solve Riemann--Hilbert problems,
  Cecotti--Vafa ({\cite{cv}}) and then Gaiotto--Moore--Neitzke ({\cite{gmn}})
  used multiple integrals quite similar to the one defining paralogarithmic
  resurgence monomials, but without a parameter.
\end{remark}

\subsection{Antipodal symmetry}

A very intriguing feature of the use of paralogarithmic resurgence monomials
for the effective solution of inverse problems is that, alongside the sought
analytic objects in $z \sim \infty$ that they enable to build, they also yield
other analytic objects, but in $z \sim 0$. This fact, which is totally absent
when using hyperlogarithms, of course has its origin in the simultaneous presence of
$z$ and of $z^{- 1}$ in the mould formulas defining the paralogarithms and
there are in fact compact formulas given in section 11.4 of {\cite{e_twist}},
relating the fundamental objects $\Theta_c (z)$ and $\Theta_c (c^2 z^{- 1})$.

That feature is reminiscent of the phenomenon of ``UV/IR mixing''
for a number of perturbative series in Physics; on the mathematical side,
anyway, it raises many interesting questions (e.g. about possible overlapping
of the analytic continuations of the 2 objects, from $\infty$ and $0$
respectively).

\section{Appendix}

\subsection{Alien derivations and related moulds}

Systems of alien operators $\Gamma_{\eta}$, notably of alien derivations
$(\Delta_{\eta})$, acting on spaces of resurgent functions which have minors
with isolated singularities $\eta_1, \eta_2, \ldots$ (we keep here the
notations of {\cite{e_twist}}: the $\eta' s$ designate the successive singular
points and the $\omega' s$ the increments: $\omega_i = \eta_i - \eta_{i - 1}$)
on a given direction $d$ in the Borel plane, can be defined either:
\begin{enumerate}
  \item by averaging various analytic continuations following the direction
  $d$ and circumventing the points $\eta_i$ by the right or by the left (with
  a tag $\varepsilon_i = +$ or $-$, resp.), involving families of weights
  $ d^{\tmscript{\left(\begin{array}{ccc}
    \varepsilon_1 & \ldots & \varepsilon_r\\
    \omega_1 & \ldots & \omega_r
  \end{array}\right)}}$
  
  \item or, as mentionned in section 2 above, by relating them to the lateral
  operators $\Delta_{\omega}^+  (\tmop{or} \Delta_{\omega}^-)$, with
  so--called ``transition moulds'' denoted by $(\tmop{led}^{\bullet})$ and
  $(\tmop{red}^{\bullet})$, respectively.
\end{enumerate}

We refer to {\cite{e1,e3,e_snag,e_twist}} for the former point of view (see
also {\cite{sau_sn}}) and focus here on the latter and recall the formula:
\[ \Gamma_{\eta_n} = \sum_{\omega_1 + \ldots + \omega_r = \omega_n}
   led^{(\omega_1, \ldots, \omega_r)} \Delta_{\omega_r}^+ \ldots
   \Delta_{\omega_1}^+ \]
\begin{proposition}
  Let and $\tmop{red}^{\bullet}$ and $\tmop{led}^{\bullet}$ be the 2 ``lateral
  moulds'' defining a family of alien operators; then we
  have({\cite{e_twist}}):
  
  The operators $\Gamma_{\eta}$ are derivations $\Leftrightarrow$
  $\tmop{red}^{\bullet}$ is an alternel mould $\Leftrightarrow$
  $\tmop{led}^{\bullet}$ is an alternel mould
\end{proposition}

\begin{definition}
  The organic alien derivations $dom$ (there is another related system, called $don$, which also satisfies exponential bounds at the arborified level; we refer to \cite{e_twist}) can be defined by the explicit
  lateral alternel moulds:
  \[ \tmop{redom}^{(\omega_1, \ldots, \omega_r)} \assign (- 1)^r  \frac{1}{2} 
     \frac{\omega_1 + \omega_r}{\omega_1 + \ldots + \omega_r} = -
     \tmop{ledom}^{(\omega_1, \ldots, \omega_r)} \]
\end{definition}

The crucial property satisfied by the organic system of alien derivation is
contained in the following:

\begin{proposition}
  The contracted arborified of the moulds $\tmop{redom}^{\bullet}$,
  $\tmop{ledom}^{\bullet}$ have exponential bounds: $\exists H, K > 0$ such
  that, for any forest $F$ with $r$ summits
  \[ | \tmop{redom}^F | \leqslant H^r \tmop{and} \quad | \tmop{ledom}^F |
     \leqslant K^r \]
\end{proposition}

It is this proposition that entails that the ordinary differential operators produced by expressing the bridge equation with the organic alien derivations have exponential growth estimates; as mentioned above, the ones corresponding to the standard system of alien derivations do not (this is related to the fact that the latter are the homogeneous components of the directional derivation which is the infinitesimal generator of a Stokes diffeomorphism and such a derivation is generically divergent \cite{e2,e3,sau_book}), which make them less convenient for the synthesis problem. Beside the organic system of derivations, they are many other ones, with interesting properties and beautiful analytic or combinatorial constructions; we refer to \cite{e_twist} for detailed information on this.

\subsection{Resurgence monomials and systems of alien derivations}

We have given in section 6 the multiple integrals that define the
paralogarithmic resurgence monomials which are orthogonal to the operators
$\Delta_{\omega}^+$ and worked with them, as in our presentation we have
chosen to take as input the operators obtained through the action of the
$\Delta_n^+$.

It is however possible to express the monomials which are orthogonal to the
alien {\tmem{derivations}} of any given system, by an average of similar
integrals formulas. Indeed, by considering the same integrands as in section
6.1, we can decide to integrate on the direction $\mathbbm{R}_{>}$
successively in the variables $y_1, \ldots, y_r$ by prescribing that, when
integrating over $y_i$, we give a $\varepsilon_i = +$ tag (resp.
$\varepsilon_i = -$) if we bypass the point $y_{i + 1}$ from below (resp. from
above); correspondingly, we shall obtain an integral that depends on the
sequence $\tmmathbf{\varepsilon}$ of the signs $\varepsilon_i$.

Now, if, for a given sequence $\omega_1, \ldots, \omega_r$ we perform an
{\tmem{average of these integrals, with the family $\tmmathbf{\delta}$ of
coefficients $\left( d^{\tmscript{\left(\begin{array}{ccc}
  \varepsilon_1 & \ldots & \varepsilon_r\\
  \omega_1 & \ldots & \omega_r
\end{array}\right)}} \right)$ used to define the corresponding alien
derivation}} ({\cite{e_twist}}, 2.3 and 6.3), we will automatically get a
family $({}^{\tmmathbf{\delta}} \mathcal{U}^{\tmmathbf{\omega}} (z))$of
monomials which are orthogonal to them, namely $\Delta$--friendly for this
particular set of derivations.

For any given family of averaging weights defining good (wrt to growth
property of the ordinary operators produced by the corresponding bridge
equation) alien derivations, the passage from the (arborified of the) ${}^+
\mathcal{U} (z)$ to the (arborified of the) ${}^{\tmmathbf{\delta}}
\mathcal{U} (z)$ can be expressed by a product by a scalar (ar)mould with
exponential bounds and thus, if we are able to prove that the ${}^+
\mathcal{U}^{\omega_1 \ldots \omega_r} (z)$ satisfy exponential growth
estimates, then we shall have for free the same property for monomials
$\mathcal{U}$ that correspond to a good family of alien derivations, e. g. the
organic ones.

It is {\tmem{most valuable in the proofs}}, if we wish to formulate the
synthesis problem when we take as input bridge equation operators obtained
through the action of good alien derivations: we can still focus on the
monomials ${}^+ \mathcal{U}^{\omega_1 \ldots \omega_r} (z)$ and the mould
combinatorics takes care of the rest.

  \vspace{1cm}

Fr{\'e}d{\'e}ric Fauvet, IRMA, Universit{\'e} de Strasbourg et CNRS, 7 rue Descartes

67084 \ Strasbourg Cedex, France. 

{\small\tt fauvet@math.unistra.fr}

\end{document}